\documentclass[12pt]{article}
\usepackage{a4wide}
\usepackage{amsmath}
\usepackage{amssymb}
\usepackage{amsopn}
\usepackage{graphicx}
\begin{document}
{\renewcommand{\thefootnote}{\fnsymbol{footnote}}
\hfill  AEI--2004--049\\ 
\medskip
\hfill gr--qc/0407018\\
\medskip
\begin{center}
{\LARGE  The Volume Operator\\ in Spherically Symmetric Quantum Geometry}\\
\vspace{1.5em}
Martin Bojowald\footnote{e-mail address: {\tt mabo@aei.mpg.de}}
and Rafal Swiderski\footnote{e-mail address: {\tt swiderski@aei.mpg.de}}
\\
\vspace{0.5em}
Max-Planck-Institut f\"ur Gravitationsphysik, Albert-Einstein-Institut,\\
Am M\"uhlenberg 1, D-14476 Golm, Germany
\vspace{1.5em}
\end{center}
}

\setcounter{footnote}{0}

\newtheorem{lemma}{Lemma}

\newcommand{\proofend}{\raisebox{1.3mm}{\fbox{\begin{minipage}[b][0cm][b]{0cm}
\end{minipage}}}}
\newenvironment{proof}{\noindent{\it Proof:} }{\mbox{}\hfill \proofend\\\mbox{}}

\newcommand{\case}[2]{{\textstyle \frac{#1}{#2}}}
\newcommand{\lP}{\ell_{\mathrm P}}

\newcommand{\md}{{\mathrm{d}}}
\newcommand{\Kern}{\mathop{\mathrm{ker}}}
\newcommand{\tr}{\mathop{\mathrm{tr}}}
\newcommand{\sgn}{\mathop{\mathrm{sgn}}}
\newcommand{\Lam}{\Lambda}
\newcommand{\vt}{\vartheta}
\newcommand{\vp}{\varphi}

\newcommand*{\R}{{\mathbb R}}
\newcommand*{\N}{{\mathbb N}}
\newcommand*{\Z}{{\mathbb Z}}
\newcommand*{\Q}{{\mathbb Q}}
\newcommand*{\C}{{\mathbb C}}

\begin{abstract}
 The spherically symmetric volume operator is discussed and all its
 eigenstates and eigenvalues are computed. Even though the operator is
 more complicated than its homogeneous analog, the spectra are
 related in the sense that the larger spherically symmetric volume
 spectrum adds fine structure to the homogeneous spectrum. The
 formulas of this paper complete the derivation of an explicit
 calculus for spherically symmetric models which is needed for future
 physical investigations.
\end{abstract}

\section{Introduction}

In complicated theories such as general relativity or its possible
quantization, loop quantum gravity \cite{Rev}, much information can
usually be gleaned by studying simpler situations in symmetric
contexts. Classically one thus obtains special exact solutions, which
are relevant if perturbations around them are stable. In the quantum
theory, symmetric models always present an approximation, whose
relevance can be investigated by comparison with less symmetric
situations. Examples for exact symmetric solutions were found early
on, with applications in cosmology (isotropic space,
\cite{Static,Friedmann}) and black holes (spherical symmetry,
\cite{Schwarzschild}). Quantum gravity in the Wheeler--DeWitt form
has, similarly, first been studied in isotropic models \cite{DeWitt}
followed later by inhomogeneous ones \cite{K:EinsteinRosen}.

In loop quantum gravity there is a systematic procedure to derive
symmetric models and to relate them to the full theory
\cite{SymmRed,PhD}. Also here, applications were first done in the
simplest, isotropic context followed by anisotropic
models. Applications include a general non-singular behavior in those
models \cite{Sing,IsoCosmo,HomCosmo,Spin}, new information about
initial conditions \cite{In}, and several phenomenological scenarios
\cite{Inflation,Bounce,NonChaos}.

Spherical symmetry presents an inhomogeneous model with infinitely
many kinematical degrees of freedom, but also interesting physical
applications for black holes. At the same time, results from
homogeneous models can be checked by comparing with a more
complicated, inhomogeneous context. In many respects, spherical
symmetry yields the simplest inhomogeneous model. Classically, there
are only finitely many physical degrees of freedom and the system is
exactly soluble \cite{SphKl,Kuchar}. The loop quantization
also presents some features that can be expected to simplify
calculations, such as the existence of a flux representation
\cite{SphSymm} which was helpful in homogeneous models. We will see
here that also the volume spectrum can be obtained explicitly which is
essential for calculations involving, e.g., the Hamiltonian
constraint. However, the volume will already be more complicated than
in homogeneous models, which will appear as special cases.

One of the aims of studying different symmetric models is to weaken
the degree of symmetry step by step, in order to check the robustness
of results.  Here, our focus will be on the effect of level splitting
(of volume eigenvalues; for the area operator see \cite{AreaOp}) which
is a consequence of weakening the symmetry and familiar from atomic or
molecular spectroscopy. A difference in the case of quantum geometry
is that the symmetry can only be reduced in discrete steps rather than
by a continuous deformation. Mathematically this is implied by the
fact that symmetric states are not ordinary states but distributions
in a less symmetric model. Yet, the effect is visible, e.g.\ when
comparing isotropy with anisotropy. It will be seen similarly, but in
a much more finely structured way, when comparing homogeneity with
inhomogeneity, even when only a single vertex is considered.

This effect illustrates properties of the volume and may also shed
light on the full case. Further applications, which will be studied
elsewhere, include, e.g., the Hamiltonian constraint, observables,
horizons in quantum geometry. The calculations will be more cumbersome
than in homogeneous models since the volume operator has eigenstates
different from those of flux operators, and the transformation between
the eigenbases is quite involved. We will present here some of the
formulas which can be helpful for explicit investigations.

\section{Spherical symmetry}

The spherically symmetric model from the point of view of quantum
geometry has been described in \cite{SphSymm}, where its states and
basic operators have been derived from a symmetry reduction. Here we
collect the information relevant for studying the volume operator.

\subsection{Classical reduction}

Spherically symmetric connections and densitized triads on a manifold
$\Sigma\cong B\times S^2$ can be written as
\begin{equation} \label{A}
 A=A_x(x)\Lam_3\md r+(A_1(x)\Lam_1+A_2(x)\Lam_2)\md\vt+
(A_1(x)\Lam_2-A_2(x)\Lam_1)\sin\vt\md\vp+ \Lam_3\cos\vt\md\vp
\end{equation}
and
\begin{equation} \label{E}
 E=E^x(x)\Lam_3\sin\vt\frac{\partial}{\partial x}+
(E^1(x)\Lam_1+E^2(x)\Lam_2)\sin\vt\frac{\partial}{\partial\vt}+
(E^1(x)\Lam_2-E^2(x)\Lam_1)\frac{\partial}{\partial\vp}
\end{equation}
with real functions $A_x$, $A_1$, $A_2$, $E^x$, $E^1$ and $E^2$ on a
one-dimensional, radial manifold $B$ with coordinate $x$ (in addition
to polar coordinates $\vt$ and $\vp$). The ${\rm su}(2)$-matrices
$\Lam_I$ are constant and identical to $\tau_I=-\frac{i}{2}\sigma_I$
(with Pauli matrices $\sigma_I$) or a rigid rotation thereof. The
symplectic structure is given by
\begin{equation}
 \Omega_{B}=\frac{1}{2\gamma G}\int_B\md x(\md
A_x\wedge\md E^x+ 2\md A_1\wedge\md E^1+2\md A_2\wedge\md E^2)
\end{equation}
with the gravitational constant $G$ and the Barbero--Immirzi parameter
$\gamma$.

The reduced theory has residual U(1) gauge transformations
corresponding to rotations around $\Lam_3$ and generated by the Gauss
constraint
\begin{equation}
 G[\lambda]=\int_B\md x\lambda (E^x{}'+2A_1E^2-2A_2E^1)\approx0 \,.
\end{equation}
The components $(A_1,A_2)$
and $(E^1,E^2)$ transform in the defining SO(2)-representation, such
that the components
\begin{eqnarray}
 A_{\vp}(x) &:=& \sqrt{A_1(x)^2+A_2(x)^2}\,,\\
 E_{\vp}(x) &:=& \sqrt{E^1(x)^2+E^2(x)^2}
\end{eqnarray}
are gauge invariant. The corresponding internal directions
\begin{eqnarray}
 \Lambda_{\vp}^A(x) &:=& (A_1(x)\Lam_2-A_2(x)\Lam_1)/A_{\vp}(x)\,,\\
 \Lambda_{\vp}^E(x) &:=& (E^1(x)\Lam_2-E^2(x)\Lam_1)/E^{\vp}(x)
\end{eqnarray}
are rotated simultaneously such that
\begin{equation}
 \cos\alpha(x):=\Lambda_{\vp}^A(x)\cdot\Lambda_{\vp}^E(x)
\end{equation}
is also gauge invariant. Parameterizing
\begin{eqnarray}
 \Lambda_{\vp}^A(x) &=:& \Lam_1\cos\beta(x)+\Lam_2\sin\beta(x)\,,\\
 \Lambda_{\vp}^E(x) &=:& \Lam_1\cos\left(\alpha(x)+\beta(x)\right)+
\Lam_2\sin\left(\alpha(x)+\beta(x)\right)
\end{eqnarray}
introduces the angle $\beta(x)$ which is pure gauge.

In these variables which take into account the gauge structure we have
the symplectic form
\begin{eqnarray}
 \Omega_{B} &=& \frac{1}{2\gamma G}\int_B\md x \left(\md
A_x\wedge\md E^x+2\md A_{\vp}\wedge\md (E^{\vp}\cos\alpha)+
2\md\beta\wedge \md(A_{\vp}E^{\vp}\sin\alpha)\right)\nonumber\\
 &=& \frac{1}{2\gamma G}\int_B\md x \left(\md A_x\wedge\md E^x+ \md
A_{\vp}\wedge\md P^{\vp}+ \md\beta\wedge\md P^{\beta}\right)
\end{eqnarray}
with momenta
\begin{equation} \label{Pphi}
 P^{\vp}(x):=2E^{\vp}(x)\cos\alpha(x)
\end{equation}
conjugate to $A_{\vp}$ and
\begin{equation} \label{Pbeta}
 P^{\beta}(x):=2A_{\vp}(x)E^{\vp}(x)\sin\alpha(x)=
A_{\vp}(x)P^{\vp}(x)\tan\alpha(x)
\end{equation}
conjugate to $\beta$. The Gauss constraint then takes the form
\begin{equation} \label{Gauss}
 G[\lambda]=\int_B\md x\lambda(E^x{}'-P^{\beta})\approx0
\end{equation}
which is easily solved by $P^{\beta}=E^x{}'$ while the function
$A_x+\beta'$ is manifestly gauge invariant.

Thus, in this model flux variables (i.e., momenta of the connection),
are not identical to triad components. In particular, some metric
components in
\begin{equation} \label{metric}
 \md s^2=E^x(x)^{-1}E^{\vp}(x)^2\md x^2+E^x(x)(\md\vt^2+\sin^2\vt\md\vp^2)\,,
\end{equation}
and in particular the volume $V({\cal I})=4\pi \int_{\cal I} \md x
\sqrt{|E^x|} E^{\vp}$ of the shell ${\cal I}\times S^2\subset\Sigma$
given by an interval ${\cal I}\subset B$, are rather complicated
functions of the momenta as well as of connection components.

\subsection{Loop quantization}

States in the connection representation depend on the components
$A_x$, $A_{\vp}$ and $\beta$. The loop representation is based on
cylindrical states with respect to graphs $g$ in the one-dimensional
radial manifold $B$, consisting of a collection of non-overlapping
edges $e$ whose endpoints define the vertex set $V(g)$. We assume a
given orientation of $B$ and choose the same one for all edges. Each
vertex $v\in V(g)$ then has one outgoing edge $e^+(v)$ and one
incoming edge $e^-(v)$. An orthonormal basis of states is given by
spin network states
\begin{equation} \label{GaugeInvSpinNetwork}
 T_{g,k,\mu}=\prod_{e\in g} \exp\left(\tfrac{1}{2}i k_e
\smallint_e(A_x+\beta')\md x\right)  \prod_{v\in V(g)}
\exp(i\mu_v A_{\vp(v)})
\end{equation}
in the gauge invariant form, with labels $k_e\in\Z$ for
U(1)-representations on all edges $e\subset B$ and $\mu_v\in\R$ for
all vertices $v\in B$. Thus, in addition to ordinary holonomies of
$A_x$ along edges there are vertices with point holonomies
$\exp(i\mu_v A_{\vp(v)})$ which can be thought of as representing
orbital edges in the symmetry orbits $S^2$. Gauge invariance requires
only that the differences $k_{e^+(v)}-k_{e^-(v)}$ for any two edges
$e^+(v)$ and $e^-(v)$ meeting in a common vertex $v$ must be
even. Otherwise the edge and vertex labels are free.

Connection components are quantized via holonomies and act directly by
multiplication. For instance, looking only at a single vertex $v$ and
suppressing all labels but $\mu_v$, we obtain
\begin{eqnarray}
 \sin(\mu A_{\vp})T_{\mu_v} &=& -\tfrac{1}{2}i
(T_{\mu_v+\mu}-T_{\mu_v-\mu}) \label{sinFlux}\\
 \cos(\mu A_{\vp})T_{\mu_v} &=& \tfrac{1}{2}
(T_{\mu_v+\mu}+T_{\mu_v-\mu}) \label{cosFlux}
\end{eqnarray}
as basic operators representing the transversal connection component
$A_{\vp}$ which takes values in the Bohr compactification
$\bar{\R}_{\rm Bohr}$ of the real line.  Flux components act as
\begin{eqnarray}
 \hat{E}^x(x) T_{g,k,\mu} &=& \frac{\gamma\lP^2}{8\pi}
\frac{k_{e^+(x)}+k_{e^-(x)}}{2} T_{g,k,\mu} \label{Exspec}\\
 \int_{\cal I}\hat{P}^{\vp}T_{g,k,\mu} &=& \frac{\gamma\lP^2}{4\pi}
\sum_{v\in{\cal I}\cap V(g)} \mu_v T_{g,k,\mu}\label{Ppspec}\\
 \int_{\cal I}\hat{P}^{\beta} T_{g,k,\mu} &=&
 \frac{\gamma\lP^2}{4\pi}\sum_{v\in{\cal I}\cap V(g)} k_v T_{g,k,\mu}
 \label{Pbspec}
\end{eqnarray}
with $k_v:=\frac{1}{2}(k_{e^+(v)}-k_{e^-(v)})\in\Z$, which shows that
they all have discrete spectra (normalizable eigenstates). 

The momenta $P^{\vp}$ and $P^{\beta}$ transform as scalars of density
weight one and thus have to be integrated over intervals ${\cal
I}\subset B$ in order to yield well-defined flux operators.
Alternatively, they can be regarded as operator valued distributions
\[
 \hat{P}^{\vp}(x) T_{g,k,\mu} = \frac{\gamma\lP^2}{4\pi} \sum_{v\in
V(g)} \mu_v \delta(x,v) T_{g,k,\mu}\,.
\]

\section{Volume operator}

Classically, the volume of a spherically symmetric region between two
spherical shells located at points $x_1, x_2\in B$ is given by
\begin{equation}
 V({\cal I})=4\pi\int_{{\cal I}}\md x \sqrt{|E^x|}E^{\vp}
\end{equation}
where ${\cal I}=[x_1,x_2]$. Since $|\hat{E}^x|$ is a non-negative
self-adjoint operator, we can directly use it in this expression. The
triad component $E^{\vp}$, on the other hand, is more complicated to
quantize since it is related to the basic variables by (\ref{Pphi})
which also involves connection components through $\alpha$. Using
(\ref{Pphi}) and (\ref{Pbeta}), we obtain
\begin{equation} \label{Ephi}
 E^{\vp} =\frac{P^{\vp}}{2\cos\alpha}=
\tfrac{1}{2}P^{\vp} \sqrt{1+\tan^2\alpha}=
\tfrac{1}{2}\sqrt{(P^{\vp})^2+A_{\vp}^{-2}(P^{\beta})^2}\,.
\end{equation}

In order to quantize $E^{\vp}$ and diagonalize the resulting operator
it suffices to look at the action on eigenstates of $\hat{E}^x$ and
thus $\hat{P}^{\beta}$ since this does not restrict the
$A_{\vp}$-dependence of states. In the expression for $E^{\vp}$ we can
then regard $P^{\beta}$ as a constant $\gamma\lP^2 k/4\pi$ and use
basic operators for $P^{\vp}$ and $A_{\vp}$. Thus, $(E^{\vp})^2$ has
the form of a classical mechanics Hamiltonian with potential
$k^2/A_{\vp}^2$. Taking this Hamiltonian at face value suggests that
the volume spectrum is continuous since the potential does not have an
absolute minimum at finite values. In fact, since $E^{\vp}$ depends on
fluxes as well as connection components there is no a priori reason to
expect a discrete spectrum. However, a continuous spectrum would be
difficult to reconcile with the full volume operator
\cite{Vol} as well as with results in homogeneous models
\cite{cosmoII,IsoCosmo,HomCosmo}.

This problem is resolved easily after noticing that there is no
operator for $A_{\vp}$ directly in a loop representation, or on the
Bohr Hilbert space in this case (see the discussion of \cite{Bohr} in
the isotropic context). Then neither the quantization of
$A_{\vp}^{-2}$ as a multiplication operator on a Schr\"odinger Hilbert
space, as understood in the above argument, is available. In fact, the
discreteness of geometric spectra depends crucially on characteristic
features of the loop representation. Following a loop quantization of
(\ref{Ephi}) and checking if the resulting spectrum will turn out to
be discrete can thus be seen as a further test of the generality of
loop results. Since the spherically symmetric volume is a rather
complicated function of the basic variables, but not basic itself, one
cannot expect a unique (up to smaller choices) quantization. But the
sensitivity of general properties like the discreteness of its
spectrum and its behavior can easily be studied with different
versions and compared with the full theory as well as other models.

In a loop quantization only holonomies of the connection are
represented as well-defined operators on the Hilbert space, not
connection components directly. This is also true for connection
components acting on a Bohr Hilbert space as in (\ref{sinFlux}),
(\ref{cosFlux}). Since the states (\ref{GaugeInvSpinNetwork}) are
orthonormal, the $\mu$-derivative at $\mu=0$ of the operators
$\sin(\mu A_{\vp})$ and $\cos(\mu A_{\vp})$ does not exist such that
an operator for $A_{\vp}$ cannot be derived in this way. Instead, we
have to use (\ref{sinFlux}) and (\ref{cosFlux}) with non-zero $\mu$ as
our basic operators in every expression where $A_{\vp}$ appears and
needs to be quantized. This in general is true for the Hamiltonian
constraint which contains curvature components of the connection. A
quantization can then be constructed by using holonomies of the
connection \cite{QSDI}, and in a homogeneous model or for orbital
components of an invariant connection point holonomies
\cite{Bohr}. Since in the spherically symmetric model the volume is
more complicated in terms of basic variables, containing in particular
connection components, we have to use the same technique here. We will
exploit this fact in order to study consequences of replacing
connection components by holonomies by means of comparing the
spherically symmetric volume spectrum with that of homogeneous models
where such a replacement is not necessary for the volume operator.

The connection components contain information about intrinsic
curvature (via the spin connection) and extrinsic curvature. In a
semiclassical regime, which we need to take into account to ensure
that the resulting volume operator will have the correct classical
limit, extrinsic curvature is small but the spin connection may not be
small when we restrict ourselves to spherical coordinates which just
leaves the freedom to transform $x$ (under arbitrary coordinate
transformations the spin connection can locally always be made as
small as desired). For our purposes it is nevertheless sufficient to
assume that the connection components, or at least $A_{\vp}$, are
small semiclassically since the spectrum will be invariant under
shifting $A_{\vp}$ by a constant in the operator used later.

We thus replace $A_{\vp}$ by $\sin (\delta A_{\vp})/\delta$ for some
$\delta\in\R\backslash\{0\}$. In the limit $\delta\to0$ we would obtain the
classical expression exactly, and for $\delta\not=0$ we have small
corrections in regions where $A_{\vp}$ is small. This leads us to
consider the Hamiltonian operator
\begin{equation} \label{H}
 \left(\frac{4\pi}{\gamma\lP^2}\right)^2
 \hat{H}_{k,\delta}=(4\pi\hat{P}^{\vp}/\gamma\lP^2)^2+\delta^2k^2
 \sin(\delta A_{\vp})^{-2}
\end{equation}
for integer $k$ and non-zero real $\delta$. It will later be used to
quantize $E^{\vp}$ and the volume. At this point we can already see
that replacing $A_{\vp}$ by the sine, which we are forced to do in the
loop representation, leads to a discrete spectrum since the potential
is now unbounded above at both sides. This fact does not depend on the
explicit realization of the replacement and is thus robust to
quantization ambiguities. In order to test the replacement we will
later use more detailed properties of the spectrum and compare it with
volume spectra in other models.

Before we continue we have to discuss the Hilbert space on which the
operator (\ref{H}) acts. Since it suffices to consider a single vertex
and we already used the eigenvalue $k$ of $\hat{P}^{\beta}$ it acts on
a single copy of the Bohr Hilbert space with $A_{\vp}\in\bar{\R}_{\rm
Bohr}$. Thus, the Hamiltonian is not yet of the usual quantum
mechanical form since so far it is not represented on the
Schr\"odinger Hilbert space. However, it is clear from (\ref{sinFlux})
that $\hat{H}_{k,\delta}$ leaves invariant any subspace spanned by
states $\exp(i(n\delta+\epsilon) A_{\vp})$, $n\in\Z$ with fixed
$\epsilon\in [0,\delta)$. Each such subspace is isomorphic
to the Schr\"odinger Hilbert space on a circle where for
$\epsilon\not=0$ we have to use a non-trivial line bundle for the
states. That the inner product on the subspace spanned by states
$\exp(i(n\delta+\epsilon) A_{\vp})$, $n\in\Z$ is identical to that for
states on a circle follows from
\begin{eqnarray*}
 \int (e^{i(n_1\delta+\epsilon)A_{\vp}})^*
 e^{i(n_2\delta+\epsilon)A_{\vp}} \md\mu_{\rm Bohr}(A_{\vp}) &=&
 \lim_{T\to\infty}\frac{1}{2T} \int_{-T}^T e^{-i(n_1-n_2)\delta
 A_{\vp}} \md A_{\vp}\\
 &=& \lim_{m\to\infty} \frac{1}{2m}\int_{-m}^m e^{-2\pi i(n_1-n_2)y}\md
y = \int_0^1 e^{-2\pi i(n_1-n_2)y}\md y\\
 &=& \delta_{n_1,n_2}
\end{eqnarray*}
using the Bohr measure and substituting $y=\delta A_{\vp}/2\pi$ and
$m=\delta T/2\pi$. Now we have an ordinary quantum mechanical system
with Hamiltonian (\ref{H}) whose spectrum can be determined with
standard methods (from this perspective, $\epsilon$ plays the role of
a $\theta$-angle).

\subsection{Spectrum of $\hat{H}$}

Using $\hat{P}_{\vp}\propto \md/\md A_{\vp}$ and substituting $\delta
A_{\vp}$ for $A_{\vp}$, the operator family (\ref{H}) labeled by
$\delta$ satisfies the identity $\hat{H}_{k,\delta}=\delta^2
\hat{H}_{k,1}$ such that it is enough to determine the spectrum for
$\delta=1$. Moreover, since the sequence of states running through
$\hat{P}_{\vp}$-eigenvalue zero is most important, we will focus on
the case $\epsilon=0$, for which states are periodic functions on a
circle. There are then standard techniques (see, e.g.,
\cite{MorseFeshbach}, p.\ 732) to determine the spectrum of the
operator
\[
 \hat{H}_{\rho}=-\frac{\md^2}{\md \xi^2}+ \frac{\rho^2-\case{1}{4}}{\sin^2\xi}
\]
where for later convenience we introduce
$\rho:=\sqrt{k^2+\frac{1}{4}}>0$ which for our case of integer $k$ is
always non-integer.

If we know an eigenstate $\psi_n^{\rho}$ with eigenvalue
$\lambda_n$ labeled by some parameter $n$, i.e.\ $\hat{H}_{\rho}
\psi_n^{\rho}=\lambda_n\psi_n^{\rho}$, we can find
eigenstates of $\hat{H}_{\rho\pm1}$ by applying the operators
\begin{equation}
 \hat{a}_{\rho}^{\pm}:= \pm \frac{\md}{\md \xi}+(\rho-\case{1}{2})\cot \xi
\end{equation}
to the state $\psi_n^{\rho}$. (For our values of $\rho$, $\rho\pm 1$
will not be of the form $\sqrt{k^2+\frac{1}{4}}$ with integer
$k$. Nevertheless, these values can be used in intermediate steps to
construct the eigenstates.) This follows from the identities
\begin{eqnarray}
 \hat{a}_{\rho}^-\hat{a}_{\rho}^+ &=&
\hat{H}_{\rho}- (\rho-\case{1}{2})^2\\
 \hat{a}_{\rho+1}^+\hat{a}_{\rho+1}^- &=& \hat{H}_{\rho}-
(\rho+\case{1}{2})^2
\end{eqnarray}
which imply
\[
 \hat{H}_{\rho-1} \hat{a}_{\rho}^+ = (\hat{a}_{\rho}^+
\hat{a}_{\rho}^- + (\rho-\case{1}{2})^2) \hat{a}_{\rho}^+ =
\hat{a}_{\rho}^+ (\hat{a}_{\rho}^- \hat{a}_{\rho}^+ +
(\rho-\case{1}{2})^2) = \hat{a}_{\rho}^+ \hat{H}_{\rho}
\]
and similarly
\[
 \hat{H}_{\rho+1}\hat{a}_{\rho+1}^- = \hat{a}_{\rho+1}^- \hat{H}_{\rho}\,.
\]
Thus, if $\psi_n^{\rho}$ is an eigenstate of $\hat{H}_{\rho}$ with
eigenvalue $\lambda_n$, then $\hat{a}_{\rho}^+\psi_n^{\rho}$ is
an eigenstate of $\hat{H}_{\rho-1}$ and
$\hat{a}_{\rho+1}^-\psi_n^{\rho}$ an eigenstate of $\hat{H}_{\rho+1}$
with the same eigenvalue $\lambda_n$.

In order to normalize the new eigenstates we compute
\begin{equation} \label{normplus}
 \int |\hat{a}_{\rho+1}^-\psi_n^{\rho}|^2 \md \xi= \int
\psi_n^{\rho *} \hat{a}_{\rho+1}^+\hat{a}_{\rho+1}^- \psi_n^{\rho}
\md \xi = (\lambda_n-(\rho+\case{1}{2})^2) \int
|\psi_n^{\rho}|^2\md \xi
\end{equation}
and
\begin{equation} \label{normminus}
 \int |\hat{a}_{\rho}^+\psi_n^{\rho}|^2 \md \xi= \int
\psi_n^{\rho *} \hat{a}_{\rho}^-\hat{a}_{\rho}^+ \psi_n^{\rho}
\md \xi = (\lambda_n-(\rho-\case{1}{2})^2) \int
|\psi_n^{\rho}|^2\md \xi\,.
\end{equation}
Since all eigenvalues of $\hat{H}_{\rho}$ are larger than the
potential minimum
$\rho^2-\frac{1}{4}=(\rho+\frac{1}{2})(\rho-\frac{1}{2})$, we must
have started with an eigenvalue $\lambda_n>(\rho-\case{1}{2})^2$. When
applying $\hat{a}_{\rho}^-$ several times, we get eigenstates to
$\hat{H}_{\rho}$ with higher and higher values for $\rho$, and since
$\lambda_n$ stays the same in this process and the integrals in
(\ref{normplus}) are non-negative, (\ref{normplus}) is consistent only
if $\lambda_n=(\rho+n+\frac{1}{2})^2$ and $n\in\N_0$ a non-negative
integer. The state with the highest allowed $\rho$ at fixed
$\lambda_n$, i.e.\ for $n=0$, is then annihilated by
$\hat{a}_{\rho+1}^-$ as a consequence of (\ref{normplus}):
\[
 \hat{a}_{\rho+1}^- \psi_{0}^{\rho}=0
\]
which has normalized solution
\begin{equation}
 \psi_0^{\rho}(\xi) =
 \frac{\sqrt{\case{1}{2}\Gamma(2\rho+2)}}{2^{\rho}\Gamma(\rho+1)}
\sin^{\rho+1/2}\xi\,.
\end{equation}

Normalized eigenstates with values $n>0$ are obtained by
applying $\hat{a}_{\rho}^+$ and using (\ref{normminus}) with
$\lambda_n-(\rho-\frac{1}{2})^2= (2\rho+n)(n+1)$ such that
$\psi_{n+1}^{\rho-1}=[(2\rho+n)(n+1)]^{-1/2} \hat{a}_{\rho}^+
\psi_n^{\rho}$. The eigenstate $\psi_n^{\rho}$ can thus be obtained
from the known $\psi_0^{\rho+n}$ as
\[
 \psi_n^{\rho}=\sqrt{\frac{\Gamma(2\rho+n+1)}{n!\,\Gamma(2\rho+2n+1)}}
\:\hat{a}_{\rho+1}^+\cdots \hat{a}_{\rho+n}^+ \psi_0^{\rho+n}\,.
\]
Using
\[
 \hat{a}_{\rho}^+ = \frac{\md}{\md \xi}+(\rho-\case{1}{2})\cot \xi =
(\sin \xi)^{1/2-\rho}\frac{\md}{\md \xi}(\sin \xi)^{\rho-1/2}= -(\sin
\xi)^{3/2-\rho}\frac{\md}{\md\cos \xi} (\sin \xi)^{\rho-1/2}
\]
such that
\[
 \hat{a}_{\rho+1}^+\cdots \hat{a}_{\rho+n}^+= (-1)^n(\sin
\xi)^{1/2-\rho}\frac{\md^n}{(\md\cos \xi)^n} (\sin \xi)^{\rho+n-1/2}
\]
this yields all eigenstates as
\begin{equation}
 \psi_n^{\rho}(\xi) = (-1)^n
\frac{\sqrt{(\rho+n+\case{1}{2})\Gamma(2\rho+n+1)}}{2^{\rho+n}\Gamma(\rho+n+1)
\sqrt{n!}} \sin^{1/2-\rho}\xi \frac{\md^n}{(\md \cos
\xi)^n} \sin^{2(\rho+n)}\xi
\end{equation}
of $\hat{H}_{\rho}$ with eigenvalue $\lambda_{n}=(\rho+n+1/2)^2$,
$n\in\N_0$. If $\rho=1/2$, i.e.\ $k=0$, the potential vanishes and the
poles at $x=0,\pi$ disappear. In this case boundary conditions leave
more freedom since states do not need to vanish at $x=0,\pi$ which
results in the additional eigenvalue $\lambda=0$ for the case
$\rho=1/2$.

These eigenfunctions can be expressed in terms of Legendre
functions\footnote{It is amusing to note that such as in rotationally
symmetric systems in quantum mechanics also in spherically symmetric
quantum geometry Legendre functions play an important role (with the
difference that here the labels are always non-integer).}
$P_{\nu}^{\mu}(z)$ (see Appendix)
\begin{equation}
 \psi_n^{\rho}(\xi)= \sqrt{(\rho+n+\case{1}{2})\Gamma(2\rho+n+1)/n!}
\,\sqrt{|\sin \xi|}\, P_{\rho+n}^{-\rho}(\cos \xi)
\end{equation}
such that the sum of degree and order, $\nu+\mu=n$, is a
non-negative integer and the order $\mu=-\rho$ is negative.

\subsection{Volume spectrum}

We now have to construct the volume operator from the basic building
blocks $\hat{E}^x$ and
$\hat{E}^{\vp}=\frac{\gamma\lP^2}{8\pi}\sqrt{\hat{H}}$. In particular,
expressions like $(P^{\vp})^2$ and $(P^{\beta})^2$ in $H$ have to be
regularized upon quantization since $\hat{P}^{\vp}$ and
$\hat{P}^{\beta}$ are operator valued distributions.

We do that by decomposing a given interval ${\cal I}$, whose volume
$V=4\pi\int_{\cal I} \md x\sqrt{|E^x|}E^{\vp}$ we want to quantize, into
subintervals ${\cal I}_i= [x_i-\epsilon/2, x_i+\epsilon/2]$ each of
size $\epsilon$: ${\cal I}=\bigcup_i {\cal I}_i$. The volume can then
be written as
\begin{eqnarray*}
 V({\cal I}) &=& 4\pi\int_{\cal I} \md x\sqrt{|E^x|}E^{\vp}= 4\pi\sum_i
 \int_{{\cal I}_i} \md x\sqrt{|E^x|}E^{\vp}\\
 &\approx& 4\pi\sum_i \sqrt{|E^x(x_i)|} E^{\vp}(x_i)\epsilon=
2\pi\sum_i \sqrt{|E^x(x_i)|} \sqrt{P^{\vp}(x_i)^2\epsilon^2+
P^{\beta}(x_i)^2\epsilon^2/A_{\vp}(x_i)^2}\\ &\approx& 2\pi
\sum_i \sqrt{|E^x(x_i)|} \sqrt{(\smallint\nolimits_{{\cal I}_i}
P^{\vp})^2+ (\smallint\nolimits_{{\cal I}_i}
P^{\beta})^2/A_{\vp}(x_i)^2}
\end{eqnarray*}
where now only squares of fluxes occur which can immediately be
quantized. We now choose the subdivision of ${\cal I}$ fine enough
such that each ${\cal I}_i$ contains at most one vertex of the spin
network state we act on, and immediately obtain the volume operator
\begin{equation}
 \hat{V}({\cal I})= \frac{\gamma\lP^2}{2} \sum_v \sqrt{|\hat{E}^x(v)|
\hat{H}(v)}
\end{equation}
where only vertices of graphs appear since otherwise $\hat{P}^{\vp}$
and $\hat{P}^{\beta}$, and thus $\hat{H}(v)=-\partial^2/\partial
A_{\vp}(v)^2- (\sin A_{\vp}(v))^{-2}\,\partial^2/\partial\beta(v)^2$,
would annihilate the state.

Using the spectra of $\hat{E}^x$ and $\hat{H}$, we find the volume
spectrum as
\begin{equation} \label{VolSpec}
 V_{k,n}= (\case{1}{2}\gamma\lP^2)^{3/2} (4\pi)^{-1/2} \sum_v
 \sqrt{\case{1}{2}|k_{e^+(v)}+k_{e^-(v)}|}
 \left(n_v+\case{1}{2}\left(1+\sqrt{1+(k_{e^+(v)}-k_{e^-(v)})^2}\right)\right)
\end{equation}
where $k_{e^+(v)}$ is the label of the edge outgoing from $v$,
$k_{e^-(v)}$ the label of the edge incoming to $v$, and $n_v$ a label
of the vertex with $n_v\in\N_0$ if $k_{e^+(v)}\not= k_{e^-(v)}$ and
$n_v\in\N_0\cup\{-1\}$ if $k_{e^+(v)}= k_{e^-(v)}$. Moreover, for the
spectrum on gauge invariant states the difference $k_{e^+(v)}-
k_{e^-(v)}$ must be even \cite{SphSymm}, and thus also the sum
$k_{e^+(v)}+ k_{e^-(v)}$.

The $A_{\vp}$-dependence of volume eigenstates is now via an
$\hat{H}$-eigenstate $\psi_{n(v)}^{\rho(v)}(A_{\vp})$, where
$\rho(v)=\frac{1}{2}\sqrt{1+(k_{e^+(v)}-k_{e^-(v)})^2}$ using the
eigenvalue $\frac{1}{2}(k_{e^+(v)}-k_{e^-(v)})$ of $\hat{P}^{\beta}$,
explicitly given by
\begin{equation}
 S_{g,k,n}=\prod_{e\in g} \exp\left(\tfrac{1}{2}i k_e
\smallint_e(A_x+\beta')\md x\right)  \prod_{v\in V(g)}
\psi_{n_v}^{\frac{1}{2}\sqrt{1+(k_{e^+(v)}-k_{e^-(v)})^2}}(A_{\vp(v)})\,.
\end{equation}

\subsection{Level splitting}

\begin{figure}
\begin{center}
 \setlength{\unitlength}{9mm} 
\begin{picture}(13,20)
 \put(0,0){\line(1,0){13}} 
 \put(0,0){\vector(0,1){20}}
 \input{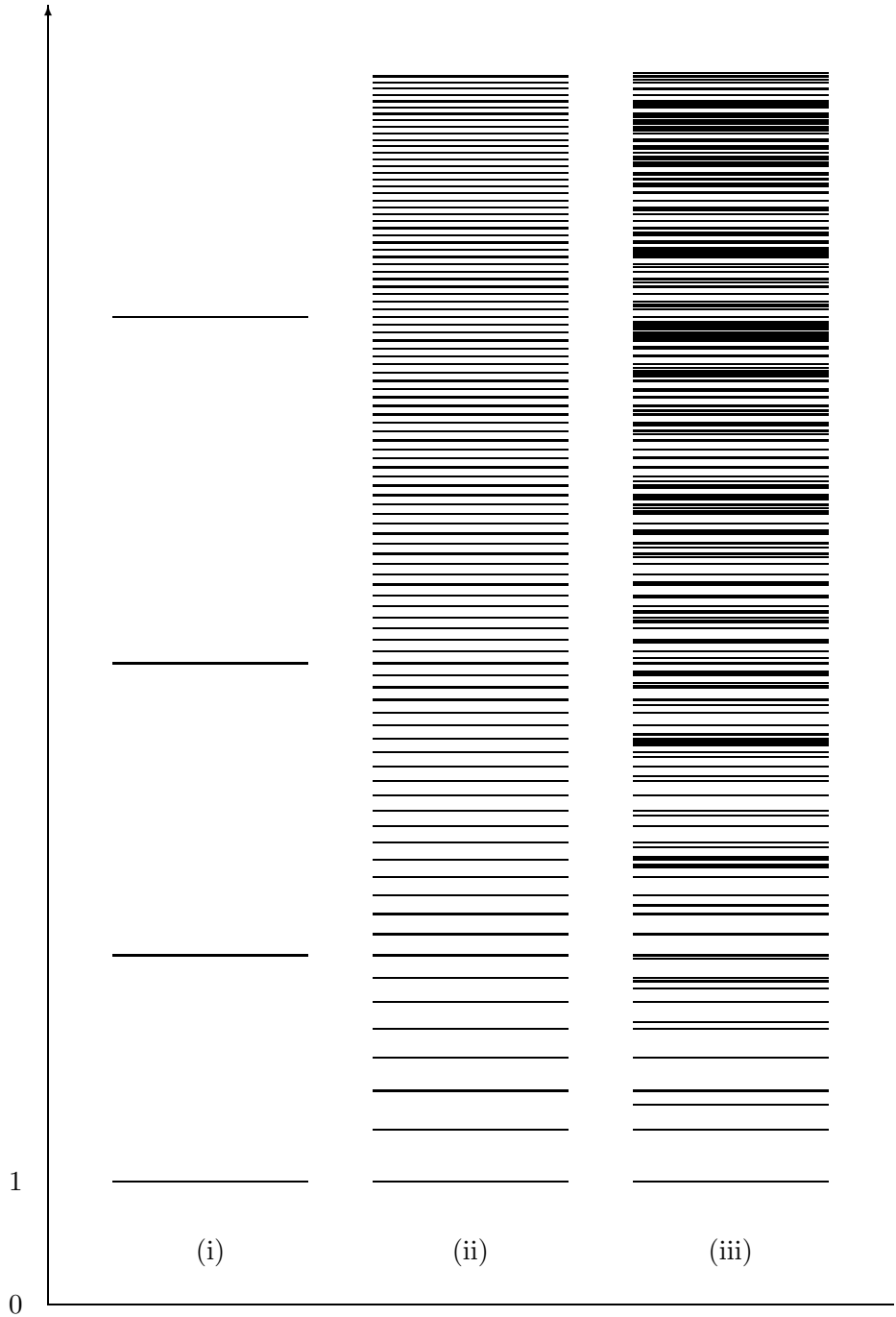}
 \put(2.5,0.8){\makebox(0,0){\small (i)}}
 \put(6.5,0.8){\makebox(0,0){\small (ii)}}
 \put(10.5,0.8){\makebox(0,0){\small (iii)}}
 \put(-0.5,0){\makebox(0,0){\small 0}}
 \put(-0.5,1.9){\makebox(0,0){\small 1}}
\end{picture}
\end{center}
 \caption{Comparison between the main series of the isotropic volume
 spectrum $n^{3/2}$, $n\in\N_0$ (i), the homogeneous LRS (single
 rotational axis) volume spectrum $m\sqrt{n}$, $m,n\in\N_0$ (ii) and
 one vertex contribution of the spherically symmetric volume spectrum
 (\ref{VolSpec}) (iii). Constant factors resulting from the
symplectic structure have been ignored. \label{Split}}
\end{figure}

As in any loop quantization, the spherically symmetric volume operator
has a discrete spectrum. Due to the symmetry, however, the set of
eigenvalues (considering also non-zero values for $\epsilon$) is the
whole positive real line (see \cite{Bohr} for a discussion of this
issue in the isotropic context; discussing observables may reduce the
spectrum to a discrete subset \cite{Velhinho}). It is nevertheless
instructive to compare the main series of the volume spectra, which
are obtained by restricting to a separable subspace corresponding to
equidistant flux eigenvalues containing zero, for different symmetric
models. Similarly to the spectroscopy of atoms, one then observes that
levels split when the symmetry is weakened \cite{AreaOp}, as
illustrated in Fig.~\ref{Split}.

\begin{figure}
 \includegraphics[width=14cm,keepaspectratio]{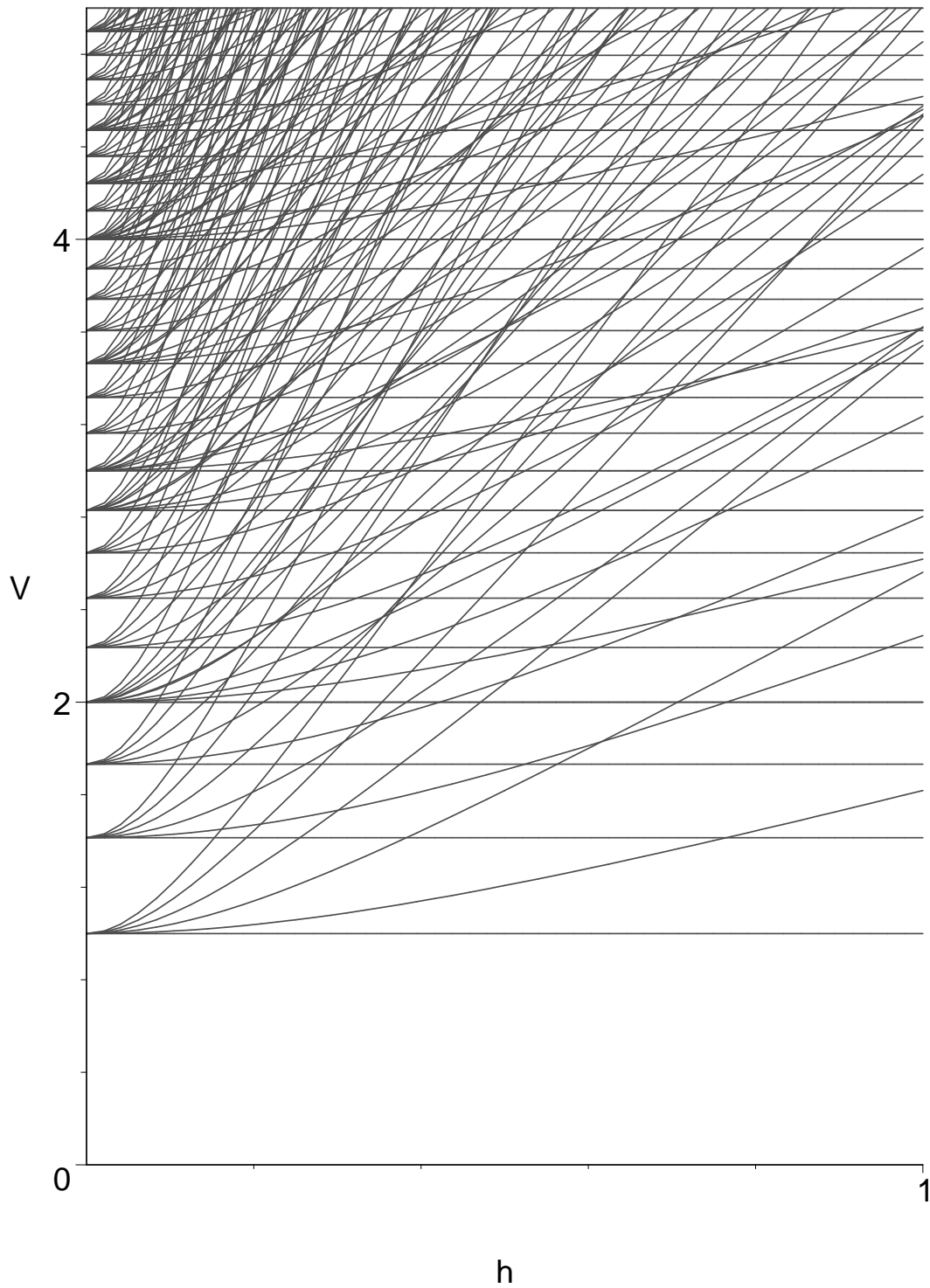}
 \caption{Explicit splitting of the LRS levels with an auxiliary
parameter $0<h<1$. \label{SplitLow}}
\end{figure}

Comparing in particular the homogeneous LRS spectrum with a single
spherically symmetric vertex contribution shows that the spectra
differ only in details such that the homogeneous levels split only
slightly. Moreover, all the levels of the LRS spectrum are preserved
and reappear in the spherically symmetric spectrum for
$k_{e^+(v)}=k_{e^-(v)}$. Since the classical geometry in a single
point of a spherically symmetric space is given by a homogeneous LRS
metric we would in fact expect a close relationship between the
quantum geometries in a given vertex. A spherically symmetric vertex
is closest to a homogeneous state if $k_{e^+(v)}=k_{e^-(v)}$ (a
homogeneous state \cite{cosmoI,HomCosmo} would be built with point
holonomies also of $A_x$, which can be seen as holonomies for closed
edges leaving and entering in the vertex which requires equal spin
numbers) in which case the spectra are identical. If $k_{e^+(v)}\not=
k_{e^-(v)}$, inhomogeneity becomes noticeable and the levels
split. One can view $k_{e^+(v)}-k_{e^-(v)}\in2\Z$ as a new quantum
number of spherically symmetric vertices characterizing the degree of
inhomogeneity. The larger $k_{e^+(v)}-k_{e^-(v)}$, the more strongly
the levels split. Even though there is no smooth transition between a
homogeneous LRS state and a spherically symmetric one since a more
symmetric state is distributional in the less symmetric model, one can
visualize the splitting by introducing a continuous parameter
$h\in[0,1]$. With this parameter, we have a family of spectra
\[
  V^{(h)}_{k_+,k_-,n}=
 \sqrt{\case{1}{2}|k_++k_-|}
 \left(n+\case{1}{2}\left(1+\sqrt{1+h^2(k_+-k_-)^2}\right)\right)\,.
\]
For $h=0$ we get the homogeneous spectrum, while $h=1$ gives all
levels of the spherically symmetric vertex contribution. Values in
between do not correspond to physical models but are introduced for
the sake of illustration. This clearly shows that levels with $k_+-k_-=0$
do not split, whereas all other levels split approximately linearly
and the stronger the larger $k_+-k_-$ is. Fig.~\ref{SplitLow} shows the
splitting of all levels which will be smaller than five after the
splitting is completed, from which one can recognize the lower part of
Fig.~\ref{Split}.

Thus, even though the spherically symmetric volume operator at
first sight looks very different from the homogeneous one (where
fluxes and triad components are identical and holonomies are not used
for the operator), its single vertex spectrum displays only slight
traces of the inhomogeneity. As discussed, there are several
ambiguities affecting the spherically symmetric operator, mainly
coming from the replacement of $A_{\vp}$. If we choose a value for
$\delta$ different from $\delta=1$ as used above, the spectrum of
$\hat{H}$ is rescaled by $\delta^2$ and that of $\hat{V}$ by
$\delta$. For this value, we would be considering a different series
of states given by $e^{in\delta A_{\vp}}$, which would similarly
define a different series of states in the LRS Hilbert space. On those
states, also the LRS volume operator is rescaled by exactly the same
value $\delta$ such that the relationship between the spectra is
preserved. If we choose a different function other than
$\delta^{-1}\sin\delta A_{\vp}$, on the other hand, the spherically
symmetric spectrum would change without a corresponding change in the
LRS spectrum. This would have an effect on the exact position of the
new levels in the spherically symmetric spectrum compared to the LRS
one.

Comparing all three cases of Fig.~\ref{Split} shows that the
difference between the isotropic spectrum and the homogeneous LRS
spectrum is much more pronounced than that between the homogeneous LRS
spectrum and the single vertex spherically symmetric spectrum. When
breaking isotropy, many more lines emerge which cannot even be
attributed to individual level splitting, and a spectrum with
increasing level distance turns into one with decreasing
distance. When breaking the homogeneity of the LRS model, on the other
hand, the single vertex spectrum results from small level splitting,
which can be considered as just fine structure of the overall
spectrum. This may be taken as an indication that the transition from
homogeneous to inhomogeneous models, though technically very involved,
is less severe in geometrical and physical consequences than the
transition from isotropy to anisotropy. Since physical results such as
the non-singular behavior \cite{Sing} have been extended from the
isotropic context to the anisotropic one
\cite{HomCosmo,Spin}, an extension to inhomogeneous models may be
feasible, too. 

Note, however, that we used only a single vertex contribution of the
spherically symmetric volume spectrum for this comparison. Since the
spectrum (\ref{VolSpec}) is a vertex sum over non-integer
contributions, it will become dense much more rapidly at large
eigenvalues than just a single vertex contribution.

\subsection{Triad spectrum}

Eq.\ (\ref{Ephi}) shows that the eigenvalues of the triad component
$E^{\vp}$ in a single vertex are given by the square root of those of
the operator $\hat{H}$,
\[
 E^{\vp}_{k,n} = \frac{\gamma\lP^2}{8\pi} \sum_v
 \left(n_v+\case{1}{2}+\case{1}{2}\sqrt{1+(k_{e^+(v)}-k_{e^-(v)})^2}
 \right)\,.
\]
Unlike for the triad component $E^x$ or flux operators, this spectrum
is not equidistant since vertex contributions are in general
irrational. Thus, for large eigenvalues the behavior is very different
and the spectrum crowds rapidly if several vertices are
considered. This observation may be of relevance for the semiclassical
limit.

It is possible to choose a quantization resulting in an equidistant
spectrum even for the triad component $E^{\vp}$. To do so, we replace
$(\hat{P}^{\beta})^2$ in $\hat{E}^{\vp}$ by
$(\hat{P}^{\beta})^2-\gamma^2\lP^4/64\pi^2$ such that the label $\rho$
used in the derivation of the $\hat{H}$-spectrum would now be
identical to the integer $k$ labeling
$\hat{P}^{\beta}$-eigenstates. Such a replacement would not spoil the
classical limit since the added constant vanishes when
$\hbar\to0$. Vertex sums such as in the $E^{\vp}$-spectrum would then
be over half-integers, which would not crowd even for many
vertices. Moreover, a vertex contribution to the spherically symmetric
volume spectrum would be identical to the LRS spectrum such that all
fine structure would disappear.

However, this replacement only works on states with non-zero
$\hat{P}^{\beta}$-eigenvalue, for otherwise the potential would be
negative and unbounded from below when $k=0$. One can avoid this by
choosing $(\hat{P}^{\beta})^2-\gamma^2\lP^4/64\pi^2$ on
$\hat{P}^{\beta}$-eigenstates with non-zero eigenvalue, and the previous
prescription $(\hat{P}^{\beta})^2$ on the zero eigenstate. Since this
special treatment of the zero eigenstate seems artificial, we do not
use the resulting operator here and regard the non-equidistant triad
spectrum and the related fine structure of the volume spectrum as a
robust consequence of the loop quantization.

\section{Operators in the volume eigenbasis}

We now have the full eigenbasis of the volume operator at our disposal
which can be used for further calculations. As another consequence we
have seen that the triad operators $\hat{E}^x$ and $\int_{{\cal
I}}\hat{E}^{\vp}$ commute with each other such that a triad eigenbasis
exists (given by the volume eigenstates). This may be useful for
investigations of the Hamiltonian constraint just as in the
homogeneous case \cite{IsoCosmo,HomCosmo,Spin}. The difference to the
homogeneous case is that the triad representation is not identical to
the flux representation which was derived for spherical symmetry in
\cite{SphSymm}. For particular calculations one then has to see which
representation is easier to use, and maybe has to transform between
the triad and the flux representation. We will use the explicit form
of the triad eigenstates together with identities for Legendre
functions (see, e.g., \cite{AbramSteg}) in order to write down this
transformation and to express some operators in both the flux and the
triad representation.

\subsection{Volume eigenstates in the flux representation}

For the Legendre functions we have the trigonometric expansion
(\ref{trig}) which gives the $A_{\vp}$-dependence of volume eigenstates
for fixed $k_+$, $k_-$ and $n$, and thus fixed $\rho$ and $n$ as
\begin{eqnarray*}
 \psi_n^{\rho}(A_{\vp}) &=& 2^{1-\rho}\pi^{-1/2}
\sqrt{n!(\rho+n+\case{1}{2}) \Gamma(2\rho+n+1)} (\sin A_{\vp})^{-\rho+1/2}\\
&& \sum_{k=0}^{\infty} \left(\begin{array}{c} n+k \\ k \end{array}\right)
\frac{\Gamma(k-\rho+\case{1}{2})}{\Gamma(-\rho+\case{1}{2})
\Gamma(n+k+\rho+\case{3}{2})} \sin((n+2k+1)A_{\vp})\,.
\end{eqnarray*}
The expansion in flux eigenstates is then obtained by using
\begin{equation} \label{Taylorsin}
 (\sin A_{\vp})^{-\rho+1/2}=\sum_{k=0}^{\infty} (-1)^k
 \left(\begin{array}{c} \case{1}{4}-\case{1}{2}\rho \\ k \end{array}\right) \cos^{2k}A_{\vp}
\end{equation}
and
\[  
 \cos^{2k}A_{\vp}=\frac{1}{2^{2k}} {2k \choose k}+\sum_{j=1}^{2 k}
 \frac{1+(-1)^j}{2^{2k}} {2k \choose k-\case{1}{2}j} \cos(j A_{\vp})
\]
as well as
\[
 \cos jA_{\vp} \sin kA_{\vp} = \case{1}{2}(\sin (j+k)A_{\vp} -\sin(j-k)A_{\vp})
\]
after which collecting terms with equal frequency leads to the
expansion of $\psi_n^{\rho}(A_{\vp})$ in terms of flux eigenstates.

\subsection{Holonomies and fluxes in the triad representation}

Holonomies $\cos A_{\vp}$ and $\sin A_{\vp}$ and the flux $\md/\md
A_{\vp}$ have simple actions (\ref{cosFlux}), (\ref{sinFlux}) and
(\ref{Ppspec}) in the flux representation, but are more complicated
when acting on volume eigenstates. Still, in some cases one can make
use of some identities satisfied by Legendre functions (which for
convenience are collected in the Appendix) in order to get explicit
expressions for the action of holonomy and flux operators in the
volume eigenbasis.

The simplest operator will be $\cos A_{\vp}$ since we can use
(\ref{Recursecos}) and obtain a simple recursion where $\rho$ is
fixed:
\begin{eqnarray} \label{cosVol}
 (\cos A_{\vp}) \psi_n^{\rho}(A_{\vp})&=& \frac{1}{\sqrt{2\rho+2n+1}}\left(
\sqrt{\frac{(n+1)(2\rho+n+1)}{2\rho+2n+3}}\:
\psi_{n+1}^{\rho}(A_{\vp})\right.\\
&&\qquad+\left.\sqrt{\frac{n(2\rho+n)}{2\rho+2n-1}}\:
\psi_{n-1}^{\rho}(A_{\vp})\right)\,. \nonumber
\end{eqnarray}
If $\rho=\frac{1}{2}$, for instance, both coefficients reduce to $1/2$
and we reobtain (\ref{cosFlux}) since in this case a volume
eigenstate is a flux eigenstate.

Already $\sin A_{\vp}$ is more complicated since simple recursion
relations like (\ref{Recursesin}) would change the label $\rho$ by an
integer, which would not be a volume eigenstate (no difference between
different allowed values of $\rho=\sqrt{k^2+1/4}$ with integer $k$ is
an integer). One can use these functions in intermediate steps of the
calculation, but in the end all have to be expressed in terms of the
allowed volume eigenstates. Even when the sine has been used in
intermediate steps, there may be cancellations (either directly or
after using identities like (\ref{Recursetwo}) for Legendre functions)
in some cases such that in the end the result will be expressed as a
linear combination of finitely many volume eigenstates.  Instead of
using a recursion relation, we can use the expansion (\ref{Taylorsin})
and use the action of $\cos A_{\vp}$. This results in an infinite
linear combination of volume eigenstates.

The square $\sin^2 A_{\vp}$, on the other hand, does have a simple
action since it can simply be written as $1-\cos^2 A_{\vp}$, for
which we can use the action derived before. We then obtain
\begin{eqnarray*} 
 (\sin^2 A_{\vp}) \psi^\rho_n(A_{\vp}) &=& 2
 \frac{(\rho+n)(\rho+n+1)-1+\rho^2}{(2\rho+2n+3)(2\rho+2n-1)}
 \:\psi^\rho_n (A_{\vp}) \\ 
 & & -\frac{1}{2\rho+2n-1}
 \sqrt{\frac{n(n
 -1)(2\rho+n)(2\rho+n-1)}{(2\rho+2n+1)(2\rho+2n-3)}}\: \psi^\rho_{n-2}(A_{\vp}) \\ &
 &-\frac{1}{2\rho+2n+3}
 \sqrt{\frac{(n+1)(n+2)(2\rho+n+1)(2\rho+n+2)}{(2\rho+2n+1)(2\rho+2n+5)}}
 \:\psi^\rho_{n+2}(A_{\vp})
\end{eqnarray*}
which again for $\rho=\frac{1}{2}$ reduces to the flux result since
all coefficients collapse to $\frac{1}{4}$.

Flux operators, i.e.\ derivatives acting on the Legendre function,
would in general also change $\rho$ by an integer. When we multiply
holonomies with a flux operator, however, there may be simpler
expressions in some cases. For instance, (\ref{RecursePsin}) implies
\begin{eqnarray*} 
 \frac{\md \psi^\rho_n(A_{\vp})}{\md A_{\vp}} &=& \frac{\sqrt{2\rho+2n+1}}{2 \sqrt{1-\cos^2A_{\vp}}} \left(
 \sqrt{\frac{(n+1)(2\rho+n+1)}{2\rho+2n+3}} \psi^\rho_{n+1}
 (A_{\vp})\right. \\ 
 &&\qquad-\left.\sqrt{\frac{n(2\rho+n)}{2\rho+2n-1}}
 \psi^\rho_{n-1}(A_{\vp})\right)
\end{eqnarray*}
such that $(\sin A_{\vp}) \hat{P}^{\vp}$ would have a simple action on
volume eigenstates, even though the action of $\sin A_{\vp}$ or
$\hat{P}^{\vp}$ separately is more complicated.

Those operators are necessary to consider when composite operators
such as the Hamiltonian constraint are computed. One common example
is that of commutators between holonomies and the volume operator,
such that powers of sine and cosine act with the volume operator in
between.

\subsection{Euclidean observables}

In \cite{SphKl} the spherically symmetric system has been
reduced explicitly in complex Ashtekar variables and in particular a
complete set of observables has been found. We can use the same expressions,
translated to real variables, for the Euclidean model, which are
then given by the ADM mass
\begin{equation} \label{mass}
 m=(A_1^2+A_2^2-1)\sqrt{|E^x|} = (A_{\vp}^2-1)\sqrt{|E^x|}
\end{equation}
evaluated at the boundary at infinity, and its conjugate
\begin{equation} \label{time}
 T=(\gamma\kappa)^{-1}\int_B\md x
\frac{A_1E^1+A_2E^2}{(A_1^2+A_2^2)\sqrt{|E^x|}}=
(\gamma\kappa)^{-1}\int_B\md x \frac{P^{\vp}}{2A_{\vp}\sqrt{|E^x|}} \,.
\end{equation}
In the original variables $A_I$, $E^I$ the expressions look rather
complicated to quantize, but using the variables $A_{\vp}$ and
$P^{\vp}$ adapted to the loop quantization, we can quantize the
observables along the lines used before, after replacing $A_{\vp}$
suitably by holonomies.

We use $\sin A_{\vp}$ instead of $A_{\vp}$ in (\ref{mass}) and obtain
the operator
\begin{equation}
 \hat{m}= -\cos^2 A_{\vp}(\infty) \sqrt{|E^x(\infty)|}
\end{equation}
without ordering ambiguities. The argument $\infty$ here refers to the
boundary point of a state corresponding to spatial infinity. For
(\ref{time}) we choose the replacement $\sin A_{\vp}\cos A_{\vp}$
instead of a simple sine in order to preserve the conjugacy of the new
expressions. Moreover, here we have to be careful with the ordering
between $A_{\vp}$ and $P^{\vp}$. If we again subdivide the integration
over $B$ into integrations over intervals ${\cal I}_i$ of size
$\epsilon$ and with mid-points $x_i$, we obtain
\[
 T=(2\gamma\kappa)^{-1}\lim_{\epsilon\to0} \sum_i E^x(x_i)^{-1/2}
A_{\vp}(x_i)^{-1} \int_{{\cal I}_i}\md x P^{\vp}(x)\,.
\]
After quantizing we will be able to remove the regulator only if there
are only finitely many non-zero contributions to the sum. This is the
case if we use the ordering as indicated above, with the flux to the
right. For this ordering we obtain the operator
\begin{equation}
 \hat{T}=(2\gamma\kappa)^{-1} \sum_v \widehat{(|E^x(v)|^{-1/2})} (\sin
A_{\vp}(v)\cos A_{\vp}(v))^{-1} \hat{P}^{\vp}_v
\end{equation}
which is not symmetric in the kinematical inner product
($\hat{P}^{\vp}_v$ refers to a single vertex contribution to
(\ref{Ppspec})). If we reorder before we remove the regulator, the
resulting operator will not be well-defined since $\hat{P}^{\vp}(\sin
A_{\vp}\cos A_{\vp})^{-1}$ would have non-zero action even where the
state has no vertex. One can, however, reorder after the regulator has
been removed, such that we simply symmetrize the operator $\hat{T}$.

There are also several inverse expressions in $T$, which could make
the operator ill-defined. We indicated in the notation above that we
use a well-defined, even finite operator for $|E^x|^{-1/2}$ along the
lines of \cite{InvScale}. The inverses of $\sin A_{\vp}$ and $\cos
A_{\vp}$ are obtained as multiplication operators, which are unbounded
but densely defined. Computing the operator $\hat{T}$ explicitly is
nevertheless complicated since the flux eigenstates do not lie in the
domain of definition of the inverse sine and cosine.

On volume eigenstates the mass $\hat{m}$ has a simple action since we
only need to multiply with the cosine as in (\ref{cosVol}). For
$\hat{T}$, on the other hand, we have the flux operator which has a
complicated expression in the volume eigenbasis, and also inverse
powers of sine and cosine. The inverse sine can be directly applied to
a volume eigenstate using the relation (\ref{Recursesininv}), where
the right hand side has changed orders of the Legendre functions such
that it effectively corresponds to an infinite superposition of volume
eigenstates, but the inverse cosine again does not have the
eigenstates in its domain of definition.

Looking at the algebra of the quantum observables $\hat{m}$ and
$\hat{T}$ with a quantized Hamiltonian constraint and the issue of
self-adjointness with respect to possible physical inner products can
provide additional tests for the quantization scheme. In particular
the fact that operators like the observables and also the Hamiltonian
constraint have to be rewritten in terms of holonomies makes it more
non-trivial for the quantum algebra to reflect faithfully the
classical algebra. This is in addition to factor ordering issues that
play a role in any quantization with more complicated constraints.

\section{Discussion}

We have diagonalized the volume operator of spherically symmetric
quantum geometry explicitly, which shows that spherical symmetry can
supply a substantial amount of simplification compared to the full
theory. This makes it possible to study also composite operators
explicitly, where usually the volume operator plays a central
role. Having an explicit diagonalization of the volume operator is one
of the reasons why homogeneous models can be dealt with directly with
by now many cosmological applications. However, while explicitly
diagonalized, the spherically symmetric volume operator is much more
complicated than the homogeneous or isotropic ones. Its eigenstates
are not simply flux eigenstates but have a more complicated dependence
on the connection components via Legendre functions. Since holonomies
in general have a complicated action on volume eigenstates, composite
operators like the Hamiltonian constraint, where commutators of the
form $h[h^{-1},\hat{V}]$ between holonomies and the volume operator
appear, become more difficult to handle. Unless subtle cancellations
happen in the construction of those operators, for instance making use
of relations like (\ref{Recursetwo}) between Legendre functions, the
Hamiltonian constraint in the triad (or flux) representation will be a
difference operator of infinite order in the vertex labels $n_v$
(while the order in edge labels $k_e$ will be finite as in homogeneous
models \cite{cosmoIV,IsoCosmo,HomCosmo}). The physical significance of
this fact depends on the interpretation of solutions to the constraint
and the issue of time. Compared to the full theory the possibility of
explicit, though rather cumbersome, expressions will substantially
improve the outlook to understand the quantum system and in particular
field theoretical issues (the constraint algebra, semiclassical
states) that appear here. At least numerical computations should be
possible to implement in a straightforward way, while even this is
complicated in the full theory \cite{VolNum}.

Since the spherically symmetric volume operator is, unlike the full
one as well as homogeneous expressions, a function of the basic
expressions involving also connection components, there are more
quantization ambiguities. There are, however, no factor ordering
ambiguities and so the consequences are rather minor compared to,
e.g., quantizations for inverse powers of the volume in homogeneous
models \cite{InvScale, Ambig} where the ordering is relevant, too. In
particular the good agreement with the LRS homogeneous spectrum is
reassuring, where quantization ambiguities just affect the fine
structure of the spectrum.

\section*{Acknowledgements}

We thank Aureliano Skirzewski for discussions about the spectrum of
$\hat{H}$. Some of the work on this paper has been done at the ESI
workshop ``Gravity in two dimensions,'' September/October 2003.

\section*{Appendix}

\begin{appendix}
\renewcommand{\theequation}{\thesection.\arabic{equation}}
\setcounter{equation}{0}

\section{Properties of Legendre functions}

Recursion relations \cite{AbramSteg}:

\begin{equation} \label{Recursecos}
 zP^{\mu}_{\nu}(z)= \frac{\nu-\mu+1}{2\nu+1} P_{\nu+1}^{\mu}(z)+
\frac{\mu+\nu}{2\nu+1} P_{\nu-1}^{\mu}(z)
\end{equation}

\begin{equation} \label{Recursesin}
 \sqrt{z^2-1}P_{\nu}^{\mu}(z) = \frac{1}{2\nu+1}(P_{\nu+1}^{\mu+1}(z)-
P_{\nu-1}^{\mu+1}(z))
\end{equation}

\begin{equation} \label{Recursesincos}
 P_{\nu}^{\mu+1}(z) = (z^2-1)^{-1/2}\left( (\nu-\mu)zP_{\nu}^{\mu}(z)-
(\nu+\mu) P_{\nu-1}^{\mu}(z)\right)
\end{equation}

\begin{equation} \label{RecurseP}
 (z^2-1)\frac{\md P^{\mu}_{\nu}(z)}{\md z} = \nu z P_{\nu}^{\mu}(z)-
(\nu+\mu) P_{\nu-1}^{\mu}(z)
\end{equation}
Combining (\ref{Recursecos}) with (\ref{Recursesincos}) and then with
(\ref{Recursesin}) shows that
\begin{equation} \label{Recursetwo}
 P_{\nu+1}^{\mu+2}(z)-P_{\nu-1}^{\mu+2}(z)=
(\nu-\mu)(\nu-\mu+1)P_{\nu+1}^{\mu}(z)-
(\nu-\mu-1)(\nu+\mu)P_{\nu-1}^{\mu}(z)\,.
\end{equation}
From (\ref{Recursesincos}) together with (\ref{Recursecos}) we obtain
\begin{equation} \label{Recursesinb}
 \sqrt{z^2-1} P_{\nu}^{\mu}(z)= \frac{1}{2\nu+1}
\left((\nu-\mu+1)(\nu-\mu+2) P_{\nu+1}^{\mu-1}(z)-
(\nu+\mu)(\nu+\mu-1) P_{\nu-1}^{\mu-1}(z)\right)
\end{equation}
which is independent of the relation (\ref{Recursesin}). We can thus
combine both relations, shifting the label $\mu$ by two in
(\ref{Recursesinb}), in order to eliminate $P_{\nu-1}^{\mu+1}(z)$ on
the right hand side. After dividing by $\sqrt{z^2-1}$, we have
\begin{equation} \label{Recursesininv}
 \frac{1}{\sqrt{z^2-1}} P_{\nu}^{\mu}(z)= \frac{1}{2\mu}
\left((\nu+\mu)(\nu+\mu-1) P_{\nu-1}^{\mu-1}(z)-
P_{\nu-1}^{\mu+1}(z)\right)\,.
\end{equation}

Most of these identities involve Legendre functions with different
orders $\mu$, in particular those involving derivatives. One
combination with fixed order can be obtained by computing
\begin{eqnarray*}
 \frac{\md (1-z^2)^{1/4}P(z)}{\md \xi} &=& -\frac{\cos \xi}{2\sqrt{\sin \xi}}
P(\cos \xi)- \sin^{3/2}\xi \frac{\md P(\cos \xi)}{\md\cos \xi}\\
 &=&
\frac{z}{2(1-z^2)^{1/4}} P(z)- (1-z^2)^{3/4} \frac{\md P(z)}{\md z}
\end{eqnarray*}
where $z=\cos \xi$. Using (\ref{RecurseP}) and (\ref{Recursecos}) we
obtain
\begin{eqnarray} \label{RecursePsin}
 \frac{\md (1-z^2)^{1/4}P_{\nu}^{\mu}(z)}{\md \xi} &=& (1-z^2)^{-1/4}\left(
(\nu+\case{1}{2})z P_{\nu}^{\mu}(z)-
(\nu+\mu)P_{\nu-1}^{\mu}(z)\right)\nonumber\\
 &=& \case{1}{2} (1-z^2)^{-1/4}\left( (\nu-\mu+1) P_{\nu+1}^{\mu}(z)-
(\nu+\mu) P_{\nu-1}^{\mu}(z) \right)\,.
\end{eqnarray}

Trigonometric expansion \cite{AbramSteg}:
\begin{equation} \label{trig}
 P^\mu_\nu(\cos(\xi))= \frac{2^{\mu+1}}{\sqrt{\pi}} \sin^{\mu} \xi
 \frac{\Gamma(\nu+\mu+1)}{\Gamma(\nu+\frac{3}{2})}
 \sum_{l=0}^{\infty}\frac{(\mu+\frac{1}{2})_l (\nu+\mu+1)_l}{l! 
 (\nu+\frac{3}{2})_l}\sin\left((\nu+\mu+2l+1)\xi\right)
\end{equation}
where $(\nu)_l:=\Gamma(\nu+l)/\Gamma(\nu)$.

\section{Legendre functions for negative order and integer sum of
order and degree}

The formula
\[
 \frac{(-1)^{n+m}}{2^n n!}(1-z^2)^{m/2}
 \frac{\md^{n+m}}{\md z^{n+m}} (1-z^2)^n=P^{m}_n(z)
\]
can be found in the standard literature (e.g.,
\cite{AbramSteg,SpecialFunctions}) for positive integer labels $n$ and
$m$. This would not be applicable in our case since $\mu=-\rho$ and
$\nu=\rho+n$ are non-integer. However, for all the volume eigenstates
$\mu=-\rho$ is negative and $\mu+\nu=n$ is a non-negative integer. In
this case, the same formula applies.

\begin{lemma}
 Let $\mu<0$, and $\mu+\nu$ be a non-negative integer. Then
\begin{equation} \label{Formel} 
  P^{\mu}_\nu(z)= \frac{(-1)^{\nu+\mu}}{2^\nu \Gamma(\nu+1)}(1-z^2)^{\mu/2}
 \frac{\md^{\nu+\mu}}{\md z^{\nu+\mu}} (1-z^2)^\nu
\end{equation}
\end{lemma}

\begin{proof}
We will prove (\ref{Formel}) by induction over $n=\nu+\mu$. Since
\cite{AbramSteg}
\begin{equation} \label{Pnzero}
 P^{-\rho}_{\rho}(z)=
\frac{(z^2-1)^{\frac{1}{2}\rho}}{2^{\rho}\Gamma(\rho+1)}
\end{equation}
the identity is true for $n=0$. Let us now assume
that it is true for a given $n$. Multiplying both sides of
(\ref{Formel}) by $(1-z^2)^{-\mu/2}$ and differentiating with respect
to $z$ we obtain
\begin{eqnarray} \label{Induktionsbeweis} 
 \frac{(-1)^{\nu+\mu}}{2^\nu
 \Gamma(\nu+1)}
 \frac{\md^{\nu+\mu+1}}{\md z^{\nu+\mu+1}}(1-z^2)^\nu&=&\frac{\md}{\md z}\left(
 (1-z^2)^{-\mu/2} P^{\mu}_\nu(z)\right) \nonumber \\ &=&\mu z
 (1-z^2)^{-\mu/2-1} P^{\mu}_\nu(z)+(1-z^2)^{-\mu/2}\frac{\md
 P^{\mu}_\nu(z)}{\md z} \nonumber \\ &=&\mu z (1-z^2)^{-\mu/2-1}
 P^{\mu}_\nu(z) \nonumber \\
 &&+(1-z^2)^{-\mu/2-1}\left(-(\nu-\mu+1)P^{\mu}_{\nu+1}(z)+z
 (\nu+1)P^{\mu}_\nu(z)\right) \nonumber \\ &=& -(1-z^2)^{-(\mu+1)/2}
 P^{\mu+1}_\nu(z) 
\end{eqnarray} 
where in the last two steps use was made of 
\begin{equation} \label{Ableitungsidentitaet}
 \frac{\md P^{\mu}_\nu(z)}{\md z}=\frac{1}{1-z^2}\left((\nu+1) z
 P^{\mu}_{\nu}(z) -(\nu-\mu+1)P^{\mu}_{\nu+1}(z)\right)
\end{equation} 
and
\begin{equation} \label{ausMagnus}
 (\nu-\mu+1)P^{\mu}_{\nu+1}(z)-(\nu+\mu+1)z
 P^{\mu}_\nu(z)=\sqrt{1-z^2}P^{\mu+1}_\nu(z)\,,
\end{equation}
respectively. Here, (\ref{Ableitungsidentitaet}) follows directly from
(\ref{RecurseP}) together with (\ref{Recursecos}) whereas
(\ref{ausMagnus}) is identical to (\ref{Recursesincos}). Multiplying
equation (\ref{Induktionsbeweis}) by $-(1-z^2)^{(\mu+1)/2}$ we infer
that our identity (\ref{Formel}) is also true for $n+1$ thus
completing the proof.
\end{proof}

Again for positive integer $m$ and $n$ the function $P_n^m(z)$ (a
polynomial in this case) is normalizable with
\[
 \int_{-1}^1 P_n^m(z)^2 \md z = \frac{1}{n+\case{1}{2}}\frac{(n+m)!}{(n-m)!}\,.
\]
Also this formula holds true in our case:

\begin{lemma}
 Let $\mu<0$, and $\mu+\nu$ be a non-negative integer. Then
\begin{equation}
 \int_{-1}^1 P_{\nu}^{\mu}(z)^2 \md z =
\frac{1}{\nu+\case{1}{2}}\frac{(\nu+\mu)!}{\Gamma(\nu-\mu+1)}\,.
\end{equation}
\end{lemma}

\begin{proof}
 We use the representation (\ref{Formel}) and integrate by parts where
 boundary terms vanish:
\begin{eqnarray*}
 \int P_{\nu}^{\mu}(z)^2\md z &=& \frac{1}{2^{2\nu}\Gamma(\nu+1)^2}
\int (1-z^2)^{\mu} \left(\frac{\md^{\nu+\mu}}{\md
z^{\nu+\mu}}(1-z^2)^{\nu} \right)^2 \md z \\
 &=&\frac{-1}{2^{2\nu}\Gamma(\nu+1)^2} \int
\left(\frac{\md^{\nu+\mu-1}}{\md z^{\nu+\mu-1}}(1-z^2)^{\nu} \right)
\left((1-z^2)^{\mu} \frac{\md^{\nu+\mu+1}}{\md
z^{\nu+\mu+1}}(1-z^2)^{\nu}\right.\\
&&\left.- 2\mu z(1-z^2)^{\mu-1}
\frac{\md^{\nu+\mu}}{\md z^{\nu+\mu}}(1-z^2)^{\nu}\right) \md z\\
&=& \frac{-1}{2^{2\nu}\Gamma(\nu+1)^2} \int (1-z^2)^{(\mu-1)/2} 
\left(\frac{\md^{\nu+\mu-1}}{\md z^{\nu+\mu-1}}(1-z^2)^{\nu} \right)\\
&& \times\left((1-z^2) \frac{\md^2}{\md z^2} \left((1-z^2)^{(\mu-1)/2} 
\left(\frac{\md^{\nu+\mu-1}}{\md z^{\nu+\mu-1}}(1-z^2)^{\nu}
\right)\right)\right.\\
&&\quad- 2z(1-z^2)^{(\mu-1)/2} \frac{\md^{\nu+\mu}}{\md
z^{\nu+\mu}} (1-z^2)^{\nu}\\
&&\quad\left.+ (\mu-1)(1-(\mu-3)z^2/(1-z^2))
(1-z^2)^{(\mu-1)/2} \frac{\md^{\nu+\mu-1}}{\md z^{\nu+\mu-1}}
(1-z^2)^{\nu}\right) \md z\\
&=& \frac{-1}{2^{2\nu}\Gamma(\nu+1)^2} \int (1-z^2)^{(\mu-1)/2} 
\left(\frac{\md^{\nu+\mu-1}}{\md z^{\nu+\mu-1}}(1-z^2)^{\nu} \right)\\
&&\times
\left((1-z^2)\frac{\md^2}{\md z^2}- 2z\frac{\md}{\md
z}+\mu(\mu-1)-\frac{(\mu-1)^2}{1-z^2}\right)\\
&&\times \left((1-z^2)^{(\mu-1)/2} 
\frac{\md^{\nu+\mu-1}}{\md z^{\nu+\mu-1}}(1-z^2)^{\nu}
\right)\md z\\
&=& (\nu(\nu+1)-\mu(\mu-1))\int P_{\nu}^{\mu-1}(z)^2\md z
\end{eqnarray*}
where we used the differential equation
\begin{equation}
 (1-z^2)\frac{\md^2P_{\nu}^{\mu}(z)}{\md z^2}- 2z\frac{\md
P_{\nu}^{\mu}(z)}{\md z}+ \left(\nu(\nu+1)-
\frac{\mu^2}{1-z^2}\right)P_{\nu}^{\mu}(z)=0
\end{equation}
obeyed by the Legendre functions. Thus,
\[
 \int P_{\nu}^{\mu}(z)^2\md z = \frac{1}{(\nu-\mu)(\nu+\mu+1)} \int
P_{\nu}^{\mu+1}(z)^2 \md z
\]
and starting from
\[
 \int P_{\nu}^{-\nu}(z)^2 \md z =\frac{2}{\Gamma(2\nu+2)}
\]
which follows from (\ref{Pnzero}) the proof by induction is complete.
\end{proof}

\end{appendix}


\begin{thebibliography}{10}

\bibitem{Rev}
C.\ Rovelli,
\newblock Loop Quantum Gravity,
\newblock {\em Living Reviews in Relativity} 1 (1998) 1, [gr-qc/9710008],
\newblock http://www.livingreviews.org/Articles/Volume1/1998-1rovelli;\\
T.\ Thiemann,
\newblock Introduction to Modern Canonical Quantum General Relativity,
  [gr-qc/0110034];\\
A.\ Ashtekar and J.\ Lewandowski,
\newblock Background independent quantum gravity: A status report,
\newblock {\em Class.\ Quantum Grav.} (2004) to appear, [gr-qc/0404018]

\bibitem{Static}
A.~Einstein,
\newblock Kosmologische Betrachtungen zur allgemeinen Relativit\"atstheorie,
\newblock {\em Sitzber.\ Berlin} (1917) 142

\bibitem{Friedmann}
A.~Friedmann,
\newblock \"Uber die Kr\"ummung des Raumes,
\newblock {\em Z.\ Phys.} 10 (1922) 377--386

\bibitem{Schwarzschild}
K.~Schwarzschild,
\newblock \"Uber das Gravitationsfeld eines Massenpunktes nach der
  Einsteinschen Theorie,
\newblock {\em Sitzber.\ Deut.\ Akad.\ Wiss.\ Berlin, Phys.-Math.\ Klasse}
  (1916) 189--196,
\newblock english translation: physics/9905030

\bibitem{DeWitt}
B.~S.\ DeWitt,
\newblock Quantum Theory of Gravity. I. The Canonical Theory,
\newblock {\em Phys.\ Rev.} 160 (1967) 1113--1148

\bibitem{K:EinsteinRosen}
K.~V.\ Kucha\v{r},
\newblock Canonical Quantization of Cylindrical Gravitational Waves,
\newblock {\em Phys.\ Rev.\ D} 4 (1971) 955--986

\bibitem{SymmRed}
M.\ Bojowald and H.~A.\ Kastrup,
\newblock Symmetry Reduction for Quantized Diffeomorphism Invariant Theories of
  Connections,
\newblock {\em Class.\ Quantum Grav.} 17 (2000) 3009--3043, [hep-th/9907042]

\bibitem{PhD}
M.\ Bojowald,
\newblock {\em Quantum Geometry and Symmetry},
\newblock PhD thesis, RWTH Aachen, 2000,
\newblock published by Shaker-Verlag, Aachen

\bibitem{Sing}
M.\ Bojowald,
\newblock Absence of a Singularity in Loop Quantum Cosmology,
\newblock {\em Phys.\ Rev.\ Lett.} 86 (2001) 5227--5230, [gr-qc/0102069]

\bibitem{IsoCosmo}
M.\ Bojowald,
\newblock Isotropic Loop Quantum Cosmology,
\newblock {\em Class.\ Quantum Grav.} 19 (2002) 2717--2741, [gr-qc/0202077]

\bibitem{HomCosmo}
M.\ Bojowald,
\newblock Homogeneous loop quantum cosmology,
\newblock {\em Class.\ Quantum Grav.} 20 (2003) 2595--2615, [gr-qc/0303073]

\bibitem{Spin}
M.\ Bojowald, G.\ Date, and K.\ Vandersloot,
\newblock Homogeneous loop quantum cosmology: The role of the spin connection,
\newblock {\em Class.\ Quantum Grav.} 21 (2004) 1253--1278, [gr-qc/0311004]

\bibitem{In}
M.\ Bojowald,
\newblock Dynamical Initial Conditions in Quantum Cosmology,
\newblock {\em Phys.\ Rev.\ Lett.} 87 (2001) 121301, [gr-qc/0104072];\\
M.\ Bojowald,
\newblock Initial Conditions for a Universe,
\newblock {\em Gen.\ Rel.\ Grav.} 35 (2003) 1877--1883, [gr-qc/0305069];\\
D.\ Cartin, G.\ Khanna, and M.\ Bojowald,
\newblock Generating function techniques for loop quantum cosmology,
  [gr-qc/0405126]

\bibitem{Inflation}
M.\ Bojowald,
\newblock Inflation from quantum geometry,
\newblock {\em Phys.\ Rev.\ Lett.} 89 (2002) 261301, [gr-qc/0206054];\\
M.\ Bojowald and K.\ Vandersloot,
\newblock Loop quantum cosmology, boundary proposals, and inflation,
\newblock {\em Phys.\ Rev.\ D} 67 (2003) 124023, [gr-qc/0303072];\\
S.\ Tsujikawa, P.\ Singh, and R.\ Maartens,
\newblock Loop quantum gravity effects on inflation and the CMB
\newblock, [astro-ph/0311015];\\
M.\ Bojowald, J.~E.\ Lidsey, D.~J.\ Mulryne, P.\ Singh, and R.\ Tavakol,
\newblock Inflationary Cosmology and Quantization Ambiguities in Semi-Classical
  Loop Quantum Gravity,
\newblock {\em Phys.\ Rev.\ D} page to appear, [gr-qc/0403106]

\bibitem{Bounce}
P.\ Singh and A.\ Toporensky,
\newblock Big Crunch Avoidance in ${\rm k}=1$ Loop Quantum Cosmology,
\newblock {\em Phys.\ Rev.\ D} 69 (2004) 104008, [gr-qc/0312110];\\
J.~E.\ Lidsey, D.~J.\ Mulryne, N.~J.\ Nunes, and R.\ Tavakol,
\newblock Oscillatory Universes in Loop Quantum Cosmology and Initial
  Conditions for Inflation, [gr-qc/0406042];\\
G.~V.\ Vereshchagin,
\newblock Qualitative Approach to Semi-Classical Loop Quantum Cosmology,
  [gr-qc/0406108]

\bibitem{NonChaos}
M.\ Bojowald and G.\ Date,
\newblock Quantum suppression of the generic chaotic behavior close to
  cosmological singularities,
\newblock {\em Phys.\ Rev.\ Lett.} 92 (2004) 071302, [gr-qc/0311003];\\
M.\ Bojowald, G.\ Date, and G.~M.\ Hossain,
\newblock The Bianchi IX model in loop quantum cosmology,
\newblock {\em Class.\ Quantum Grav.} 21 (2004) 3541--3569, [gr-qc/0404039]

\bibitem{SphKl}
T.\ Thiemann and H.~A.\ Kastrup,
\newblock Canonical Quantization of Spherically Symmetric Gravity in Ashtekar's
  Self-Dual Representation,
\newblock {\em Nucl.\ Phys.\ B} 399 (1993) 211--258, [gr-qc/9310012];\\
H.~A.\ Kastrup and T.\ Thiemann,
\newblock Spherically Symmetric Gravity as a Completely Integrable System,
\newblock {\em Nucl.\ Phys.\ B} 425 (1994) 665--686, [gr-qc/9401032]

\bibitem{Kuchar}
K.~V.\ Kucha\v{r},
\newblock Geometrodynamics of Schwarzschild Black Holes,
\newblock {\em Phys.\ Rev.\ D} 50 (1994) 3961--3981

\bibitem{SphSymm}
M.\ Bojowald,
\newblock Spherically Symmetric Quantum Geometry: States and Basic Operators,
\newblock {\em Class.\ Quantum Grav.} 21 (2004) to appear, [gr-qc/0407017]

\bibitem{AreaOp}
M.\ Bojowald and H.~A.\ Kastrup,
\newblock The Area Operator in the Spherically Symmetric Sector of Loop Quantum
  Gravity,
\newblock {\em Class.\ Quantum Grav.} (2000) 3009--3043, [hep-th/9907043],
\newblock published as Section 5 of \cite{SymmRed}

\bibitem{Vol}
C.\ Rovelli and L.\ Smolin,
\newblock Discreteness of Area and Volume in Quantum Gravity,
\newblock {\em Nucl.\ Phys.\ B} 442 (1995) 593--619, [gr-qc/9411005],
\newblock Erratum: {\em Nucl.\ Phys.\ B} 456 (1995) 753;\\
A.\ Ashtekar and J.\ Lewandowski,
\newblock Quantum Theory of Geometry II: Volume Operators,
\newblock {\em Adv.\ Theor.\ Math.\ Phys.} 1 (1997) 388--429, [gr-qc/9711031]

\bibitem{cosmoII}
M.\ Bojowald,
\newblock Loop Quantum Cosmology: II. Volume Operators,
\newblock {\em Class.\ Quantum Grav.} 17 (2000) 1509--1526, [gr-qc/9910104]

\bibitem{Bohr}
A.\ Ashtekar, M.\ Bojowald, and J.\ Lewandowski,
\newblock Mathematical structure of loop quantum cosmology,
\newblock {\em Adv.\ Theor.\ Math.\ Phys.} 7 (2003) 233--268, [gr-qc/0304074]

\bibitem{QSDI}
T.\ Thiemann,
\newblock Quantum Spin Dynamics {(QSD)},
\newblock {\em Class.\ Quantum Grav.} 15 (1998) 839--873, [gr-qc/9606089]

\bibitem{MorseFeshbach}
P.~M.\ Morse and H.\ Feshbach,
\newblock {\em Methods of Theoretical Physics}, volume~1,
\newblock McGraw-Hill, New York, 1953

\bibitem{Velhinho}
J.~M.\ Velhinho,
\newblock Comments on the kinematical structure of loop quantum cosmology,
\newblock {\em Class.\ Quantum Grav.} 21 (2004) to appear, [gr-qc/0406008]

\bibitem{cosmoI}
M.\ Bojowald,
\newblock Loop Quantum Cosmology: I. Kinematics,
\newblock {\em Class.\ Quantum Grav.} 17 (2000) 1489--1508, [gr-qc/9910103]

\bibitem{AbramSteg}
M.\ Abramowitz and I.~A.\ Stegun,
\newblock {\em Pocketbook of Mathematical Functions},
\newblock Harri Deutsch, Thun, Frankfurt/Main, 1984

\bibitem{InvScale}
M.\ Bojowald,
\newblock Inverse Scale Factor in Isotropic Quantum Geometry,
\newblock {\em Phys.\ Rev.\ D} 64 (2001) 084018, [gr-qc/0105067]

\bibitem{cosmoIV}
M.\ Bojowald,
\newblock Loop Quantum Cosmology IV: Discrete Time Evolution,
\newblock {\em Class.\ Quantum Grav.} 18 (2001) 1071--1088, [gr-qc/0008053]

\bibitem{VolNum}
J.\ Brunnemann and T.\ Thiemann,
\newblock Simplification of the Spectral Analysis of the Volume Operator in
  Loop Quantum Gravity, [gr-qc/0405060]

\bibitem{Ambig}
M.\ Bojowald,
\newblock Quantization ambiguities in isotropic quantum geometry,
\newblock {\em Class.\ Quantum Grav.} 19 (2002) 5113--5130, [gr-qc/0206053]

\bibitem{SpecialFunctions}
W.\ Magnus, F.\ Oberhettinger, and R.~P.\ Soni,
\newblock {\em Formulas and Theorems for the Special Functions of Theoretical
  Physics},
\newblock Springer-Verlag, 1966

\end{thebibliography}

\end{document}